\documentclass[aps,prd,reprint,nofootinbib,longbibliography]{revtex4-1}

\usepackage{amsmath}
\usepackage{mathtools}
\usepackage{graphicx}
\usepackage[normalem]{ulem}
\usepackage[com pat=1.1.0]{tikz-feynman}
\usepackage{tikz}
\usepackage{aas_macros}
\usetikzlibrary{external}
\usepackage{comment}
\usetikzlibrary{arrows}
\usepackage[colorlinks=true, allcolors=blue]{hyperref}

\begin{document}

\title{Constructing angular distributions of neutrinos in core collapse supernova from zero-th and first moments calibrated by full Boltzmann neutrino transport}

\author{Hiroki Nagakura}
\email{hirokin@astro.princeton.edu}
\affiliation{Department of Astrophysical Sciences, Princeton University, 4 Ivy Lane, Princeton, NJ 08544, USA}
\author{Lucas Johns}
\email[]{NASA Einstein Fellow \\ ljohns@berkeley.edu}
\affiliation{Department of Physics, University of California, Berkeley, CA 94720, USA}

\begin{abstract}
Two-moment neutrino transport methods have been widely used for developing theoretical models of core-collapse supernova (CCSN), since they substantially reduce the computational burden inherent in the multi-dimensional neutrino-radiation hydrodynamical simulations. The approximation, however, comes at a price; the detailed structure of angular distribution of neutrinos is sacrificed, that is the main drawback of this approach. In this paper, we develop a novel method by which to construct angular distributions of neutrinos from the zero-th and first angular moments. In our method, the angular distribution is expressed with two quadratic functions of the neutrino angle in a piecewise fashion. We determine the best parameters in the fitting function by comparing to the neutrino data in a spherically symmetric CCSN model with full Boltzmann neutrino transport. We demonstrate the capability of our method by using our recent 2D CCSN model. We find that the essential features of the angular distributions can be well reconstructed, whereas the angular distributions of incoming neutrinos tend to have large errors that increase with flux factor ($\kappa$). This issue originates from the insensitiveness of incoming neutrinos to $\kappa$, that is an intrinsic limitation in moment methods. Based on the results of the demonstration, we assess the reliability of ELN-crossing searches with two-moment neutrino transport. This analysis is complementary to our another paper that scrutinizes the limitation of crossing searches with a few moments. We find that the systematic errors of angular distributions for incoming neutrinos lead to misjudgements of the crossing at $\kappa \gtrsim 0.5$. This casts doubt on the results of ELN-crossing searches based on two-moment methods in some previous studies.
\end{abstract}
\maketitle

\section{Introduction}\label{sec:intro}
High fidelity neutrino transport is an essential ingredient to develop theoretical models of core-collapse supernova (CCSN). The exact description of neutrino radiation field requires evolving the neutrino distribution function in six-dimensional phase space, i.e., solving Boltzmann equations. Although taking such an ab-initio approach in multi-dimensional (multi-D) CCSN models is still a significant challenge, considerable progress has been made in the last decades \cite{2008ApJ...685.1069O,2018ApJ...854..136N,2019ApJ...872..181H,2020ApJ...902..150H,2020ApJ...903...82I} (see also a recent review in \cite{2020LRCA....6....4M}). On the other hand, various approximate methods have been used in many neutrino-transport solvers, to reduce the computational cost and to avoid some intrinsic complexities in treatments of radiation-hydrodynamics with full Boltzmann neutrino transport (see e.g., \cite{1993ApJ...405..669M,2004ApJS..150..263L,2014ApJS..214...16N,2017ApJS..229...42N,2019ApJ...878..160N,2020MNRAS.496.2000C}). One of the popular approaches is the so-called two-moment method, in which the transport equation is integrated over solid angle in neutrino momentum space. The equations for the zero-th and first angular moments are solved under given a closure relation for the higher moments \cite{1972ApJ...171..127A,1981MNRAS.194..439T,2015MNRAS.453.3386J,2015PhRvD..91l4021F,2011PThPh.125.1255S,2019ApJS..241....7S,2021arXiv210202186L} (see also \cite{2020PhRvD.102h3017R} for rank-3 closure relations). Although the detailed profile of neutrino angular distributions is abandoned, it is much numerically cheaper than that solving Boltzmann equations, and it leads to a realistic solution if the closure relation can be given accurately. For these reasons, many modern three-dimensional (3D) CCSN models have been developed with moment methods \cite{2016ApJ...831...98R,2016ApJS..222...20K,2018ApJ...865...81O,2018MNRAS.477L..80K,2019MNRAS.490.4622N,2019MNRAS.482..351V,2019ApJ...873...45G,2019PhRvC.100e5802S,2020MNRAS.492.5764N,2020MNRAS.494.4665P,2020MNRAS.498L.109M,2020arXiv201010506B}.

On the other hand, the detailed information on neutrino angular distributions is mandatory to develop reliable models of CCSNe. It has been suggested that analytical closure relations fail to capture some intrinsic properties of neutrino angular distributions \cite{2017MNRAS.469.1725M,2018ApJ...854..136N,2019ApJ...872..181H,2020ApJ...903...82I}. It should be also mentioned that the full information on neutrino angular distributions is required to compute reaction kernels on some neutrino-matter interactions such as non-isoenergetic scatterings on electrons/positrons \cite{1993ApJ...410..740M,2017ApJS..229...42N}, and nucleons \cite{2020ApJ...897...43K}, and some thermal processes as pair and bremsstrahlung reactions \cite{2017ApJ...847..133R,2020PhRvD.102l3015B}. Furthermore, the detailed structure of angular distribution has been of great interest very recently in the context of fast-pairwise collective neutrino oscillations in CCSNe \cite{2005PhRvD..72d5003S,2016PhRvL.116h1101S,2016JCAP...03..042C,2017JCAP...02..019D,2017ApJ...839..132T,2017PhRvL.118b1101I,2017PhRvD..96d3016C,2018PhRvD..98j3001D,2019PhRvL.122i1101C,2019ApJ...883...80S,2019PhRvD.100d3004A,2019ApJ...886..139N,2019PhLB..790..545A,2020PhRvD.101d3009J,2020PhRvR...2a2046M,2020PhRvD.101b3018D,2020PhRvD.101f3001G,2020PhRvD.102j3017J,2020PhRvD.101d3016A,2020PhLB..80035088M,2020JCAP...05..027A,2020arXiv201206594A,2021PhRvD.103f3013C,2020arXiv201100004S,2020arXiv201101948T,2021arXiv210101278M,2021arXiv210102745R,2021arXiv210101226B,2021PhRvL.126f1302B,2021arXiv210312743S,2021arXiv210314308M,2021arXiv210315267M}. It is attributed to the fact that the so-called ELN(Electron-neutrinos Lepton Number)-crossing, in which energy-integrated angular distributions of electron-type neutrinos ($\nu_e$) and their anti-partners ($\bar{\nu}_e$) are crossing in momentum space, seems to be a crucial condition to trigger neutrino flavor conversions. Searching the ELN-crossing requires information on full angular distributions of neutrinos, implying that multi-angle neutrino transport is mandatory. As mentioned above, however, available multi-D CCSN models with full Boltzmann neutrino transport are limited by a computational burden, and most of multi-D CCSN models have been developed with approximate neutrino transport. For these reasons, alternative ways determining the possibility of fast flavor conversions are highly in demand.

Two interesting methods, by which to determine the occurrence of fast flavor conversions in the neutrino data computed with approximate neutrino transport, have been proposed in the literature. The first one is the so-called {\it zero mode search}, in which linear stability analysis of flavor conversions is performed with a condition that the wave number in the corotating frame is zero \cite{2018PhRvD..98j3001D}. In this case, the required moments in the dispersion relation are only the zero-th, first, and second ranks, indicating that two-moment methods (with a closure relation) are capable of providing the required information on the diagnosis. This method was applied to assess the possibility of fast flavor conversions in two sophisticated 3D CCSN models \cite{2019ApJ...881...36G}, and they revealed that the fast conversions likely occur in the proto-neutron star (PNS) convective layer \cite{2020PhRvD.101f3001G}. The same approximate method was also applied in simulations of another phenomena: remnants of binary neutron star mergers very recently \cite{2021arXiv210302616L}. Another novel method was proposed in \cite{2020JCAP...05..027A}; the occurrence of ELN-crossings is analyzed by introducing a new positive function, $\mathcal{F}$, which is a polynomial of directional cosines for the direction of neutrino propagation ($\mu$). The ELN angular distribution ($G(\mu)$) is integrated with $\mathcal{F}(\mu)$, the result of which can be written in terms of the angular moments of ELN. The ELN-crossing is diagnosed by comparing the sign of the two integrated quantities between with and without the weight function. In \cite{2020arXiv201206594A}, this method was applied to the same Garching 3D CCSN models \cite{2019ApJ...881...36G}, and they found positive sign of ELN-crossings in both pre-shock and post-shock regions. They also applied the same method not only $\nu_e$ and $\bar{\nu}_e$ but also heavy leptonic neutrinos ($\nu_{\mu}$, $\nu_{\tau}$, and their anti-partners) in \cite{2021PhRvD.103f3013C}, and they suggested that the occurrence of ELN-crossing is a generic feature in CCSNe.

There are, however, some caveats in their conclusions. First, both methods rely on the second moments given by an analytical closure relation, which would not be accurate enough to diagnose ELN-crossings, in particular at the semi-transparent region where neutrinos and matter are mildly coupling each other. As demonstrated in some previous studies with full Boltzmann neutrino transport \cite{2019PhRvD.100d3004A,2019ApJ...886..139N,2020PhRvD.101b3018D,2020PhRvR...2a2046M}, the occurrence of ELN-crossing is determined by a delicate competition between $\nu_e$ and $\bar{\nu}_e$ angular distributions, indicating that the uncertainty of closure relations would affect the diagnosis. Another major concern is the applicability of these approaches in the region with strongly asymmetric angular distributions. This is the region where the higher moments play a dominant role to characterize the angular distribution of neutrinos, indicating that the diagnosis with only a few moments is not reliable\footnote{In our another paper (Lucas Johns and Hiroki Nagakura in prep), we show that mis-capturing the ELN crossings occurs only by changing the treatments of higher moments. See the paper for more details.}.The crucial concern in these methods is that they are not capable of quantifying the reliability of the diagnosis by themselves, which may be only possible by making a detailed comparison with full Boltzmann neutrino transport. The computation is, however, numerically expensive and may be unfeasible in practice that the Boltzmann simulations are performed with respect to each time snapshot. This issue motivates us to develop a new approach.

In this paper, we attempt to construct angular distributions of neutrinos from the zero-th and first angular moments, and also underline how large uncertainties inherent in the approach. It should be mentioned that the full recovery of angular distribution from a few moments is, in principle, impossible from a mathematical point of view. However, we are currently considering a specific phenomena, CCSN, in which the neutrino radiation field may have specific trends or there may exist some correlations between full angular distributions and the lower moments. In this paper we scrutinize them based on the result of CCSN neutrino data computed by full Boltzmann neutrino transport \cite{2017ApJ...847..133R}. We note that, in earlier work of \cite{1992A&A...265..345J}, the detailed investigation on neutrino angular distributions has been also made based on CCSN models with Monte Carlo neutrino transport. However, the analyses were devoted to provide accurate closure relations for moment methods; hence, the results are not useful for the purpose of construction of the full angular distributions. In addition to this, they employ a CCSN model in \cite{1989A&AS...78..375J}, which is qualitatively different from the recent ones. For instances, the shock wave in their model expands rather promptly after core bounce ($\lesssim 50$ ms) even in spherical symmetry, which is not consistent with the trend found in recent CCSN simulations; the input physics was also rather old-fashioned. In this paper, we revisit the problem with one of the modern CCSN models, and develop a fitting method for which the distribution function of neutrinos ($f$) is reconstructed from the zero-th and first angular moments. It should be mentioned that we make the best parameters publicly available\footnote{The data is available from the link: \url{https://www.astro.princeton.edu/~hirokin/scripts/data.html}}, which will be useful to analyze many aspects of neutrino radiation field in CCSNe.

This paper is organized as follows. In Sec.~\ref{sec:basic}, we outline our basic strategy to construct angular distributions of neutrinos from the zero-th and first moments. In Sec.~\ref{sec:Correlations}, we make a correlation-study between some quantities in neutrino angular distributions and flux factor (the ratio of the first and zeroth angular moment; hereafter we denote it as $\kappa$), by using a result of neutrino data with full Boltzmann neutrino transport, the result of which is used to narrow down the parameter space in our method. In Sec.~\ref{sec:Construction}, we provide the detail of our method, and then compare the obtained angular distributions to those of the originals. In Sec.~\ref{sec:demo}, we demonstrate how our fitting data can be applied in multi-D CCSN models, which shows the strengths and weaknesses in our method. Based on the consideration, we discuss the applicability of our method to the ELN-crossing searches. In Sec.~\ref{sec:summary} we summarize the present study and discuss how to improve our method towards more accurate assessment of ELN-crossing. We will close the paper by providing an instruction on how to use our public available data (that capsulates best fit-parameters) to construct angular distributions of neutrinos from the zero-th and first moments.

\section{Basic strategy}\label{sec:basic}
In this section, we describe the basic strategy of our method. The detailed procedures and results will be presented in Sec.~\ref{sec:Construction}. We study angular distributions of neutrinos in post-bounce phase of CCSNe, paying particular attention to the post-shock region. In this paper, we only focus on the angular profile in $\mu$-space, i.e., azimuthal-averaged distribution in momentum space. Although the study of full 2D angular distributions is necessary for completeness, the azimuthal-averaged one is still informative to consider essential characteristics of neutrino angular profile. We postpone the study of the non-axisymmetric features in future work.

As a reference of neutrino radiation field, we employ our public available CCSN neutrino data presented in \cite{2017ApJ...847..133R}, 1D-4x model, in which the neutrino transport is determined by solving Boltzmann equations in spherical symmetry and in steady-state under a frozen fluid background given by \cite{2017ApJS..229...42N}. We note that the steady-state condition with respect to neutrino radiation field is a reasonable approximation at $t \gtrsim 100$ms after core bounce, which we focus on in this paper. The spherically symmetric condition also does not compromise the purpose of this study, since we consider the azimuthal-averaged distributions. On the other hand, there is a noticeable advantages in the model; the neutrino transport is solved with high resolutions in momentum space (see \cite{2017ApJ...847..133R} for more details). This is a beneficial property to capture detailed features of neutrino angular distributions accurately in particular at regions where the angular distributions are forward-peaked.

We start with defining a normalized distribution function $f_{\rm n}$ as
\begin{equation}
f_{\rm n} (\mu) = \frac{ f_{\rm ax}(\mu) }{  f(\mu=1)   }.
\label{eq:def_fnorm1}
\end{equation}
where $f_{\rm ax}$ denotes the azimuthal-averaged angular distribution of neutrinos, i.e.,
\begin{equation}
f_{\rm ax} (\mu) = \frac{ 1  }{ 2 \pi } \int_{0}^{2 \pi} d \phi_{\nu} f(\mu,\phi_{\nu}).
\label{eq:def_fax}
\end{equation}
where $f$ denotes the neutrino distribution function. As can be seen in Eq.~\ref{eq:def_fnorm1}, $f_{\rm n}$ corresponds to the distribution function normalized by that of neutrinos propagating outward along the radial direction. We note that the flux factor ($\kappa$) is not influenced by the normalization. It should also be mentioned that the normalization enables us to compare the angular profile of all neutrinos (arbitrary energy and species at different spatial locations) on an equal footing. In our method, we first develop a fitting method with respect to $f_{\rm n}$ at given $\kappa$. The free parameters are calibrated by the neutrino data of Boltzmann simulations\footnote{We refer angular distributions of neutrinos from the Boltzmann data for different energy ($5$ MeV to $50$ MeV, which covers the typical energy range in the semi-transparent region between PNS and the stagnated shock wave) and different species of neutrinos ($\nu_e$, $\bar{\nu}_e$ and other heavy leptonic neutrinos) in post-shock region.}. We will capsulate the obtained best-fit parameters into a table. By using the table, one can obtain angular distributions of $f_{\rm n}$ (and also $f_{\rm ax}$) from arbitrary zero-th and first-angular moments.

\begin{figure}
    \includegraphics[width=\linewidth]{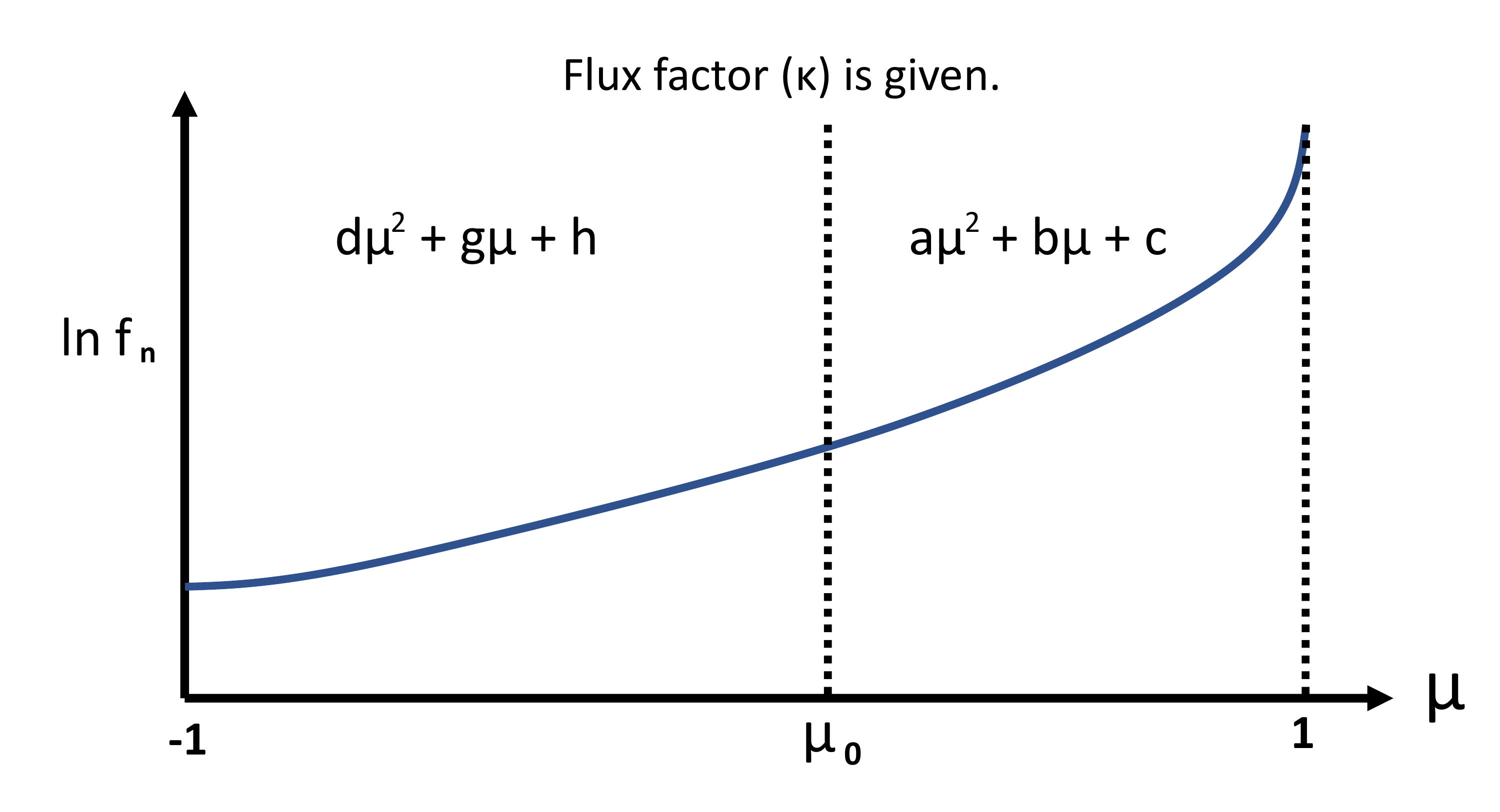}
    \caption{Schematic picture of the fitting function for angular distributions of $f_{\rm n}$. We divide the angular ($\mu$-) space into the two regions, and fit the natural logarithm of $f_{\rm n}$ with a quadratic function in a piecewise fashion. The two quadratic functions are not completely independent each other, since the angular distribution is continuously connected at $\mu=\mu_0$. See text for more details.}
    \label{fig:ReconstructBaseForm}
\end{figure}

The most straightforward fitting-formula may be a polynomial of $\mu$ or the orthogonal functions such as Legendre polynomials. Their best-fit coefficients in the polynomial for given $\kappa$ can be determined so as to minimize the errors from the collected angular distributions of $f_{\rm n}$ with the same $\kappa$ in the Boltzmann simulation. This approach is, however, not appropriate for CCSN neutrinos, since the required degree of polynomials hinges on the asymmetry of neutrinos. This leads to the need of unaffordable number of polynomials to capture the strongly forward-peaked angular distributions, i.e., the resultant fitting data will be enormous. This suggests that more appropriate fitting methods need to be considered.

One may wonder if employing specific functional forms alleviates the problem (see e.g., \cite{1992A&A...265..345J}). It should be noted, however, that the feature of the angular structure is strongly restricted, which may lead to large systematic errors. We, hence, do not employ such a specific functional form, but rather adopt a polynomial fitting approach. It should be mentioned that we avoid enormous degrees of polynomials by adopting a piecewise-fitting prescription. The essence of the fitting function is summarized in Fig.~\ref{fig:ReconstructBaseForm}. We separate the $\mu$-space into the two regions by $\mu_0$, and fit the natural logarithm of $f_{\rm n}$ with a quadratic function at each region. The two quadratic functions are not independent each other, since we impose a condition that they are continuously connected at $\mu_0$. We note that $\mu_0$ is also a free parameter in our method, which also depends on $\kappa$. The best-fit parameters are searched so as to minimize the error of reconstructed angular distribution of $f_{\rm n}$ from those in the Boltzmann simulation (see below for more details). As we shall see in Sec.~\ref{sec:Construction}, the piecewise fitting method is capable of reproducing the result of the Boltzmann simulation with the affordable number of parameters even in cases with strongly forward-peaked angular distributions ($\kappa \sim 1$).

In the fitting, we have seven parameters, $a, b, c, d, g, h$, and $\mu_0$ (see Fig.~\ref{fig:ReconstructBaseForm}), which need to be determined to each given $\kappa$. In the following sections, we describe how to determine them by using the neutrino data of Boltzmann simulation. Before moving on to the detail, we take a bit detour to a correlation-study between angular distribution of $f_{\rm n}$ and $\kappa$ in the next section. As we shall show in Sec.~\ref{sec:Construction}, the result of this correlation-study plays an important role to narrow down the parameter space in our method.

\section{Correlations}\label{sec:Correlations}

\begin{figure}
    \includegraphics[width=\linewidth]{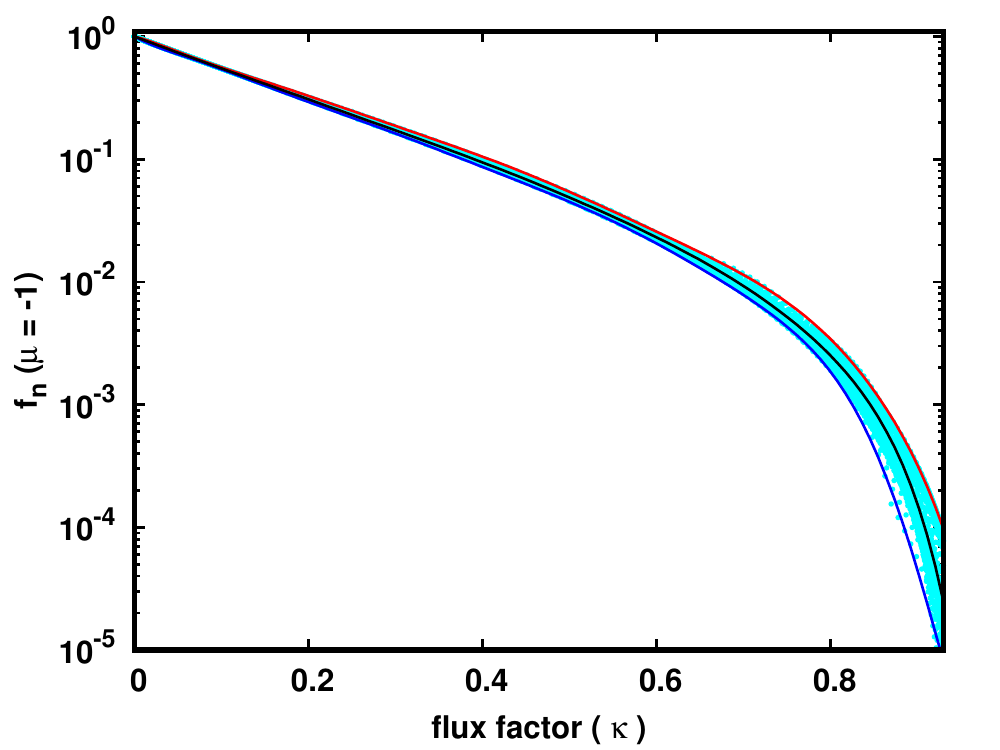}
    \caption{Cyan-dots represent $f_{\rm n}$ at $\mu=-1$ of our CCSN neutrino data computed by full Boltzmann neutrino transport (1D-4x model in \cite{2017ApJ...847..133R}), which are displayed as a function of flux factor ($\kappa$). The black line represents the best polynomial fitting function with respect to the natural logarithm of $f_{\rm n}$. The coefficients are summarized in Tab~\ref{tab:fminfitkappa}. The region surrounded by red and blue lines corresponds to the dispersion of the correlation, and we use them in the discussion of the reliability for ELN-crossing searches based on two-moment neutrino transport (see Sec.~\ref{sec:demo} for more details).
}
    \label{fig:graph_corre_kappa_fnmin1}
\end{figure}

The angular profile of neutrinos is dictated by the interplay between neutrino-matter interaction and neutrino advection. Neutrinos interact with matter through neutral- and charged current reactions during the propagation, which isotropize neutrino angular distributions. On the contrary, neutrino advection in space breaks the symmetry, in which the shape of asymmetry depends on the spatial geometry of the matter distribution. We note that the matter distribution of CCSNe is roughly spherically symmetric and the baryon density decreases monotonically with radius. By virtue of the simple geometry, the neutrino radiation field  monotonically changes from isotropic to forward-peaked distributions towards low optical depth, and the transition can be characterized solely by $\kappa$. This fact motivates us to search correlated quantities in angular distributions of $f_{\rm n}$ to $\kappa$.

There are three quantities catching our attention. The first one is $f_{\rm n}$ at $\mu=-1$. We compute them in the neutrino data of Boltzmann simulations, which are displayed as a function of $\kappa$ in Fig.~\ref{fig:graph_corre_kappa_fnmin1}. We fit the $f_{\rm n}$ at $\mu=-1$ by a decic function of $\kappa$; the best-fit function is plotted with a black line in the same figure. More specifically, the fitting function can be expressed as
\begin{equation}
\ln f_{\rm n} |_{\mu=-1} (\kappa) = \sum_{s=0}^{10} a_{s} \kappa^{s},
\label{eq:Corbetweendlogfndmuat1tokappa1}
\end{equation}
where the coefficients $a_{s}$ are provided in Table~\ref{tab:fminfitkappa}. We find that $f_{\rm n}$ at $\mu=-1$ has a very strong correlation to $\kappa$ in the region of $\kappa \lesssim 0.5$. This is attributed to the fact that the first moment is determined by the competition between the populations of outgoing- and incoming neutrinos in these opaque regions, indicating that $\kappa$ is sensitive to the angular profile for incoming neutrinos. At $\kappa \gtrsim 0.5$, however, the correlation becomes weak with increasing $\kappa$; indeed, the red and blue lines (which represent the approximate fitting functions for the upper and lower range of the correlation, respectively) deviate from the black one. This is attributed to the fact that the first moment at large $\kappa$ is characterized only by outgoing neutrinos. This is natural, since the neutrino angular distribution is forward-peaked. The uncertainty of $f_{\rm n}$ at $\mu=-1$ is, however, not a crucial issue to capture the essential characteristics of overall distribution. On the contrary, it is a crucial one to assess the ELN-crossing, the detail of which will be discussed in Secs.~\ref{sec:Construction}~and~\ref{sec:demo}.

\begin{table}[t]
\caption{The best-fit coefficients (decic function) capturing the correlation between $f_{\rm n}$ at $\mu=-1$ with $\kappa$. We note that the fitting is carried out with respect to the natural logarithm of $f_{\rm n}$ (see the text for more details).
}
\begin{tabular}{cccc} \hline
~~rank~~ & ~~value~~ & ~~rank~~ & ~~value~~ \\
 \hline \hline
$0$ & $0$  & $6$ & $-2.200 \times 10^4$ \\
$1$ & $-5.964$   &  $7$ &  $3.880 \times 10^4$  \\
$2$ & $-7.926$    & $8$ & $-4.101 \times 10^4$ \\
$3$ & $1.767 \times 10^2$   & $9$ & $2.388 \times 10^4$ \\
$4$ & $-1.572 \times 10^3$   & $10$ & $-5.901 \times 10^3$ \\
$5$ & $7.619 \times 10^3$   & - & - \\
 \hline
\end{tabular}
\label{tab:fminfitkappa}
\end{table}

Next, let us pay attention to the angular ($\mu$-direction) gradient of $f_{\rm n}$ at $\mu=-1$, since it potentially has a correlation to $\kappa$ in optically thick region. In the region of small $\kappa$, the neutrino angular distribution is nearly isotropic, implying that the gradient of $f_{\rm n}$ should be small. In the region of $\kappa \sim 1$, on the other hand, the incoming neutrinos are determined by cumulative neutrino emissions and back-scattered neutrinos during the neutrino-travel from the outside. This suggests that the angular dependence of incoming neutrinos is mild compared to the outgoing neutrinos (forward-peaked distributions). It should also be pointed out that the gradient has less influence on characterizing the angular distribution at large $\kappa$ (see Sec.~\ref{sec:Construction} for more details), indicating that it can be chosen rather arbitrarily as long as it is much smaller than that in the outgoing direction.

\begin{figure}
    \includegraphics[width=\linewidth]{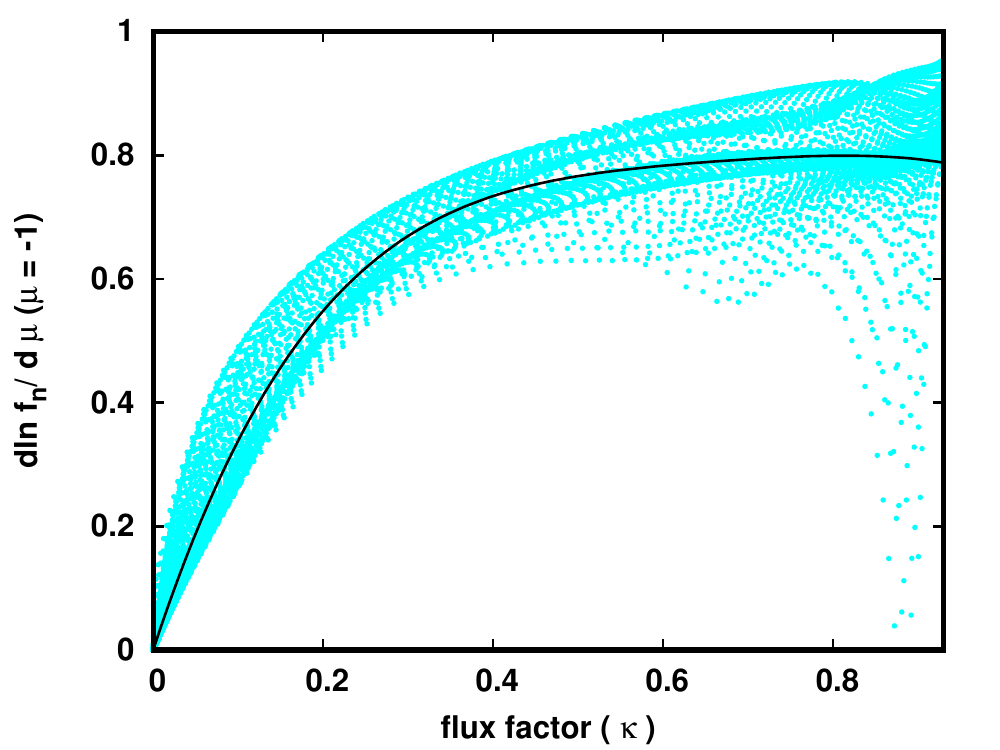}
    \caption{Same as Fig.~\ref{fig:graph_corre_kappa_fnmin1} but for the angular gradient of $\ln f_{\rm n}$ at $\mu=-1$. The fitting coefficients are summarized in Tab.~\ref{tab:dlogfdxmin1_fitcoef}.}
    \label{fig:graph_kappa_dlogfdmuatmin1}
\end{figure}

\begin{table}[t]
\caption{The fitting coefficients (quartic function of $\kappa$) to portray the correlation between the gradient of $d \ln f_{\rm n}/d \mu$ at $\mu=-1$ and $\kappa$.
}
\begin{tabular}{cr} \hline
~~rank~~ & ~~value \\
 \hline \hline
$0$ & $0$  \\
$1$ & $4.179$  \\
$2$ & $-8.771$  \\
$3$ & $8.554$  \\
$4$ & $-3.199$  \\
 \hline
\end{tabular}
\label{tab:dlogfdxmin1_fitcoef}
\end{table}

Similar as Fig.~\ref{fig:graph_corre_kappa_fnmin1}, we compute the angular gradient of $f_{\rm n}$ at $\mu=-1$ in our neutrino data of Boltzmann simulation. The result is displayed as a function of $\kappa$ in Fig.~\ref{fig:graph_kappa_dlogfdmuatmin1} (cyan dots). As expected, the correlation to $\kappa$ is robust at $\kappa \sim 0$; the gradient monotonically increases with $\kappa$. We also find that it is saturated at $\kappa \sim 0.5$, and then be almost flat profile. Although the dispersion is increased at $\kappa \gtrsim 0.5$, we confirm that the gradient is much smaller than the average; hence, it is not an issue for the retrieval of angular distributions from $\kappa$. For convenience, we fit the data by quartic function of $\mu$; the result is shown as a solid black line in Fig.~\ref{fig:graph_kappa_dlogfdmuatmin1}, and the coefficients are summarized in Table~\ref{tab:dlogfdxmin1_fitcoef}.

\begin{figure}
    \includegraphics[width=\linewidth]{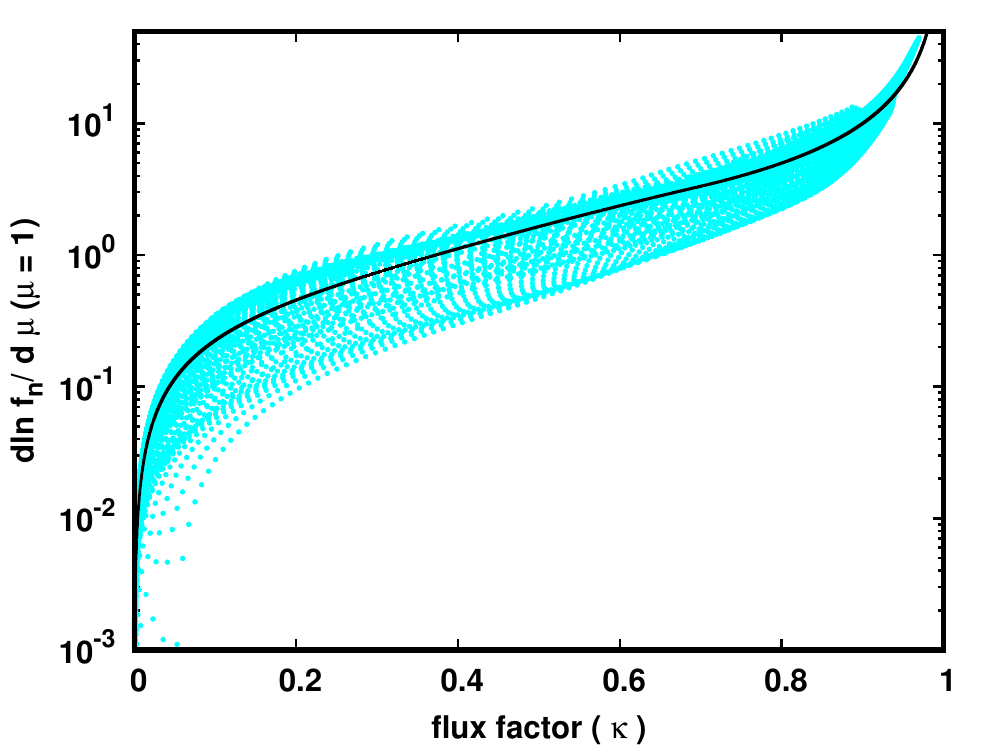}
    \caption{Same as Fig.~\ref{fig:graph_kappa_dlogfdmuatmin1} but for the angular gradient of $\ln f_{\rm n}$ at $\mu=1$. Eq.~\ref{eq:fitbykappa_dlogfdx_atx1} corresponding to the fitting function, and the best-fit one is displayed with a black line in this figure. The best-fit parameters ($u, v, w,$ and $\kappa_m$) are summarized in Tabl.~\ref{tab:dlogfdx1_fitcoef}.}
    \label{fig:graph_kappa_dlogfdmuat1}
\end{figure}

Finally, we investigate on the gradient of $f_{\rm n}$ at $\mu=1$, since the correlation to $\kappa$ can be naturally expected at $\kappa \sim 0$ and $1$. In the region of $\kappa \sim 0$, the gradient should be nearly zero, the reason of which is the same as that for the gradient at $\mu=-1$. For $\kappa \sim 1$, on the other hand, the gradient can be analytically estimated as follows. We assume that the sharp forward-peaked angular distribution is expressed with an exponential function,
\begin{equation}
f_n = \exp \{ q (\mu-1) \} \hspace{3mm} (q\gg1),
\label{eq:Corbetweendlogfndmuat1tokappa1}
\end{equation}
where $q$ corresponds to the gradient with respect to the natural logarithm of $f_{\rm n}$ at $\mu=1$. We note that the simple exponential function is not appropriate for incoming neutrinos. However, this issue can be neglected in this discussion, since the incoming neutrinos have less influence on characterizing the gradient of $f_{\rm n}$ at $\mu=1$. By virtue of the analytic expression of $f_{\rm n}$, we can express $q$ in terms of $\kappa$. By taking the limit of $q\gg1$, it can be approximately written as,
\begin{equation}
q \sim \frac{1}{1-\kappa}.
\label{eq:Corbetweendlogfndmuat1tokappa2}
\end{equation}
This consideration leads to an important conclusion that pure polynomial functions are not appropriate to fit the correlation between the gradient and $\kappa$, since the gradient diverges to infinity as $\kappa$ approaches unity. We, hence, develop a hybrid fitting function which combines a cubic function with Eq.~\ref{eq:Corbetweendlogfndmuat1tokappa2} at $\kappa \sim 1$,
\begin{equation}
\frac{d}{dx} \ln f_{\rm n} |_{\mu=1} (\kappa) =
\begin{cases}
u \kappa^3 + v \kappa^2 + w \kappa  & (\kappa \leq \kappa_m)\\
\frac{1}{1-\kappa} & (\kappa \geq \kappa_m)
\end{cases}
,
\label{eq:fitbykappa_dlogfdx_atx1}
\end{equation}
where $\kappa_m$ is the junction point for the two functions. In Eq.~\ref{eq:fitbykappa_dlogfdx_atx1}, we have four free parameters: $u, v, w,$ and $\kappa_m$, which are determined as follows. First, we impose two conditions at $\kappa_m$; the zero-th and first derivatives with respect to $\kappa$ for the two functions are continuous at $\kappa_m$. By using the conditions, we can express $u$ and $v$ as a function of $w$ and $\kappa_m$. Next, we determine $w$ independently from other parameters; we carry out root mean square fitting of Boltzmann data at the small $\kappa$, in which we fit the data with the region of $0 \le \kappa \le 0.05$, and adopt the obtained coefficient as $w$. This guarantees that fitting reproduces the Boltzmann result in the region of small $\kappa$. The rest of free parameters is $\kappa_m$, which is determined so as to minimize the error (summing up the absolute difference of the gradient between the fitting and the original.) The best-fit parameters are summarized in Table~\ref{tab:dlogfdx1_fitcoef} and the resultant fitting function is displayed in Fig.~\ref{fig:graph_kappa_dlogfdmuat1}. We confirm that it captures the essential feature of the correlation between the gradient of $f_{\rm n}$ at $\mu=1$ and $\kappa$.

\begin{table}[t]
\caption{The best-fit parameters in Eq.~\ref{eq:fitbykappa_dlogfdx_atx1} (for the gradient of $f_{\rm n}$ at $\mu=1$) to the result of Boltzmann simulation.
}
\begin{tabular}{cc} \hline
~~coefficients~~ & ~~best-fit \\
 \hline \hline
$u$ & $7.883$  \\
$v$ & $-2.124$  \\
$w$ & $2.3859$  \\
$\kappa_m$ & $0.6975$  \\
 \hline
\end{tabular}
\label{tab:dlogfdx1_fitcoef}
\end{table}

With taking advantage of the three correlations presented in this section, we develop a method to construct the full angular distribution of $f_{\rm n}$ from $\kappa$. The detail is described in the next section.

\section{Construction of angular distribution of $f_{\rm ax}$ from $\kappa$}\label{sec:Construction}
Let us turn our attention to searching the best-fit parameters in our method, for which it captures the essential characteristics of angular distribution of $f_{\rm n}$ obtained in Boltzmann simulation (see Fig.~\ref{fig:ReconstructBaseForm}). We note that this search requires a detailed exploration in seven dimensional parameter space at each $\kappa$, indicating that it is computationally expensive. We alleviate the computational burden by using the three correlations described in Sec.~\ref{sec:Correlations}. Below, we spell out explicitly our parameter-search method optimized by the correlations.

In the region of $\mu \le \mu_0$, we determine values of the three coefficients in the quadratic function: $d, g,$ and $h$; thus, three independent conditions need to be imposed. The two of them can be given from the correlations described in the previous section. The first one is $f_{\rm n}$ at $\mu=-1$ and the other is its angular ($\mu$-direction) gradient at the same angular point. Using the polynominal fit describing each correlation to $\kappa$ (the coefficients are given in Tables~\ref{tab:fminfitkappa}~and~\ref{tab:dlogfdxmin1_fitcoef} for the former and the latter, respectively), they can be determined independently and solely from $\kappa$. We can, hence, complete the determination of the three coefficients by adding one more independent condition, which is given at $\mu=\mu_0$ (see below).

In the region of $\mu \ge \mu_0$, on the other hand, we use another correlation discussed in Sec.~\ref{sec:Correlations}, the angular ($\mu$-direction) gradient of $f_{\rm n}$ at $\mu=1$, to determine values of the coefficients, $a, b$, and $c$, in the quadratic function (see Fig.~\ref{fig:ReconstructBaseForm}). The correlation is expressed with Eq.~\ref{eq:fitbykappa_dlogfdx_atx1}; hence, the gradient can be computed solely from $\kappa$. In addition to this, $f_{\rm n}$ at $\mu=1$ is constrained to be 1 from the definition (see Eq.~\ref{eq:def_fnorm1}). Hence, as similar to the case with $\mu \le \mu_0$, we can determine the three parameters by giving one more condition.

The matching condition for the two quadratic functions at $\mu = \mu_0$ provides the rest of required conditions at both regions. In other words, the angular distribution of $f_{\rm n}$ can be fully determined by giving $\mu_0$ and $f_{\rm n} (\mu_0)$. We search the best combination of the two parameters, for which the obtained $f_{\rm n}$ becomes similar to those in the Boltzmann simulation.

With setting a $\kappa$-grid\footnote{We set up a uniform grid of $1000$ covering from $0 < \kappa \leq 0.93$. From $0.93 < \kappa < 1$, we set another uniform grid of $999$. $\kappa=0.93$ corresponds to the threshold flux factor that neutrino data of our Boltzmann simulation is available (see the text for more details).}, we at first collect $f_{\rm n}$ from the neutrino data of Boltzmann simulation, having the same value of $\kappa$. We note that flux factors extracted from the Boltzmann simulation are, in general, not on the midpoint of $\kappa$ grid; hence, we linearly interpolate them from the close spatial (radial) points. We then compute the average of collected $f_{\rm n}$ ($f_{\rm n}^{\rm Bave}$) at each $\mu$-cell used in the simulation. Next, we set trial values of $\mu_0$ and $f_{\rm n} (\mu_0)$. As mentioned, this determines the all seven parameters uniquely, i.e., the angular distribution of $f_{\rm n}$ ca be fully reconstructed, albeit temporally. We then compute the difference of the reconstructed $f_{\rm n}$ from $f_{\rm n}^{\rm Bave}$ at each $\mu$-cell. We search the combination of $\mu_0$ and $f_{\rm n} (\mu_0)$, for which the total difference (summation of the absolute difference on each $\mu$-cell) becomes the smallest.

As shown above, the dimension of our parameter-search can be reduced to two, owing to the three correlated quantities to $\kappa$. It is much less computational expensive than the original seven dimensional search; indeed, we can perform the search on a laptop. Below, the detail of our searching method for $\mu_0$ and $f_{\rm n} (\mu_0)$ is described, which seems to be efficient and stable. First, it is naturally expected that the best-fit $f_{\rm n}(\mu_0)$ would be close to $f_{\rm n}^{\rm Bave} (\mu_0)$. It is, hence, enough to search only in the vicinity of $f_{\rm n}^{\rm Bave} (\mu_0)$; in other words, $f_{\rm n}(\mu_0)$ is almost determined by giving $\mu_0$. We also accelerate and stabilize our method by imposing a reasonable condition: the best-fit $\mu_0$ and $f_{\rm n}(\mu_0)$ smoothly change with $\kappa$. This condition further narrows down the parameter space; indeed, the best $f_{\rm n}(\mu_0)$ can be easily guessed from that at nearby $\kappa$, which substantially accelerate the convergence of the search.

We note that the latter condition may result in capturing the local minimum, indicating that there may be better parameters to reflect the characteristics of $f_{\rm n}$ in the Boltzmann simulation. In fact, there is no guarantee that the parameters for the global minimum distribute continuously with $\kappa$. It should be stressed, however, that the missing of the global minimum does not compromise the accuracy of our method, since the dispersion of the angular distributions found in Boltzmann simulations overwhelm the discrepancy between the local and global one (see below for more details). It should also be pointed out that, this condition makes our method stable. This is because, if we discard the latter condition, the best combination of $\mu_0$ and $f_{\rm n}(\mu_0)$ would vary discontinuously with a small change of $\kappa$, and the obtained angular spectrum of $f_{\rm n}$ results in large fluctuations with $\kappa$. Furthermore, the smooth change of all coefficients to $\kappa$ is convenient for making a data-table, in which the best-fit parameters (and other useful quantities) for each $\kappa$ are capsulated. When we employ the table for CCSN simulations or other relevant analyses, we will construct the angular distribution of neutrinos with arbitrary $\kappa$. This indicates that we need interpolations of the angular distributions from those at the nearest $\kappa$-cell on both sides. The discrete change of the angular distribution, however, may lead to unphysical outcome. The smooth condition, hence, plays an important role as suppressing these artifacts.

There still remain two more things before completing our method. First, we slightly correct the fitting parameters (except for $\mu_0$) in the two quadratic functions so as to ensure that the flux factor computed from the obtained $f_{\rm n}$ exactly matches the given $\kappa$. In fact, we only focus on the deviation from the result of Boltzmann simulation to determine the parameters. As a result, there is not guarantee that the reconstructed $f_{\rm n}$ produces the same $\kappa$. It may be, however, no problem to neglect the discrepancy, since we find that the error is small ($< 5 \%$). This correction is just for the usability of our data-table.

We make an ad-hoc prescription to correct these coefficients, in which we change the overall shape of the angular distribution of $f_{\rm n}(\mu_0)$ by introducing a scale factor of $\eta$. It is multiplied to all six coefficients ($a, b, c, d, g,$ and $h$ in Fig.~\ref{fig:ReconstructBaseForm}); in other words, the angular distribution is tilted by $\eta$. We iteratively search the value of $\eta$ until the flux factor computed from the angular distribution of $f_{\rm n}$ coincides with given $\kappa$. The advantage of this method is that $f_{\rm n}$ at $\mu=1$ does not change, since the logarithm of $f_{\rm n}$ at this point is zero. Hence, the condition of Eq.~\ref{eq:def_fnorm1}, which will be used to retrieve $f_{\rm ax}$ from $f_{\rm n}$, is not affected by the correction. It is also deserved to be mentioned that the matching condition at $\mu_0$ is guaranteed, since the two quadratic functions are changed with the same scale factor. On the other hand, some correlated quantities to $\kappa$ (see Sec.~\ref{sec:Correlations}) are slightly influenced by the correction. However, the impact is tiny; indeed, we confirm that the resultant $f_{\rm n}$ sustains the essential trend of correlations.

As the second remark, let us describe how we determine the parameters for strongly forward-peaked angular distributions which the Boltzmann simulation does not provide. In the neutrino data, we find that the maximum flux factor is $\sim 0.95$ for our considered range of neutrino energy ($5$MeV to $50$MeV) in the post-shock region. We also find that the data sample is small around the edge of the maximum. Thus, we apply our method only to the region of $0 < \kappa \leq 0.93$. For $0.93 < \kappa$, we need to develop a prescription to determine the best-fit parameters, which is described below. First, we assume that $\mu_0$ is constant, i.e., it is set as the same value at $\kappa=0.93$, which has been already determined by our method described above. Next, we again introduce a scale factor of $\eta$. We multiply all six coefficients in the two quadratic functions at $\kappa=0.93$ by $\eta$, and iteratively search the value of $\eta$ that the resultant $f_{\rm n}$ distribution provides the same flux factor as given $\kappa$ (the value at the $\kappa$-cell). As a result, all coefficients can be determined uniquely.

Obviously, it is a makeshift prescription. However, our prescription is capable of capturing the essential characteristics of very forward-peak angular distribution. For instance, it is compatible with our argument regarding the gradient of $f_{\rm n}$ at $\mu=1$ (see Eqs.~\ref{eq:Corbetweendlogfndmuat1tokappa1}~and~\ref{eq:Corbetweendlogfndmuat1tokappa2}); indeed, we confirm that the prescription provides the consistent solution with the argument. On the contrary, our prescription is not capable of capturing the detailed angular structure for incoming neutrinos. As we have already pointed out, however, the angular distribution of $f_{\rm n}$ at $\mu < 0$ diverges with increasing $\kappa$ (see Fig.~\ref{fig:graph_corre_kappa_fnmin1} and also following discussions). This indicates that the fundamental improvement to our method is required to address the issue (see Sec.~\ref{sec:summary} for the discussion). We, hence, reckon that the makeshift prescription does not degrade the capability of our method.

\begin{figure*}
    \includegraphics[width=\linewidth]{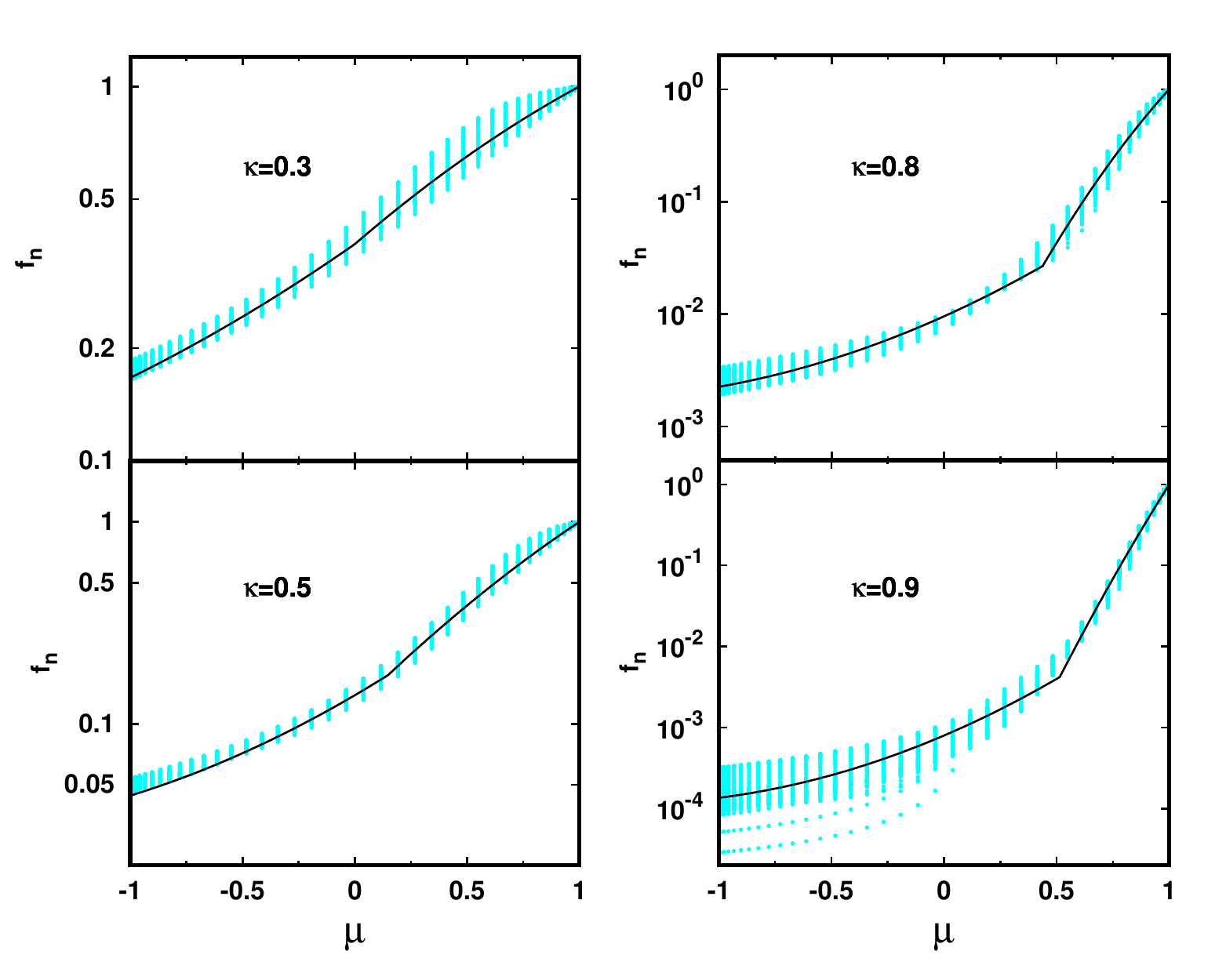}
    \caption{Black lines denotes the reconstructed angular distribution of $f_{\rm n}$ by using our method. Cyan-dots correspond to those obtained from the neutrino data of our Boltzmann simulation. We compare them in four cases: $\kappa=0.3, 0.5, 0.8,$ and $0.9$.}
    \label{fig:graph_compare_angfn_reconst_vs_Boltz}
\end{figure*}

Below, we show some essential results of reconstructed angular distribution of $f_{\rm n}$. First, we compare them to the originals in Boltzmann simulation (see Fig.~\ref{fig:graph_compare_angfn_reconst_vs_Boltz}). We select the cases with four representative $\kappa$: $\kappa=0.3, 0.5, 0.8,$ and $0.9$. As shown in these plots, the obtained $f_{\rm n}$ successfully captures the essential profile in those obtained from the CCSN simulation, which lends confidence to our method. It should be mentioned that the dispersion of angular distribution of $f_{\rm n}$ in the simulation increases with $\kappa$ (we note that the scale of the vertical axis is different in the panels), in particular at $\mu \lesssim 0$. This trend is consistent with our argument in Sec.~\ref{sec:Correlations}; $f_{\rm n} (\mu=-1)$ becomes less sensitive to $\kappa$ at $\kappa \sim 1$. Our result suggests that the major drawback in constructing full angular distributions from zero-th and first moments is to determine the angular profile for incoming neutrinos at high $\kappa$.

\begin{figure*}
    \includegraphics[width=\linewidth]{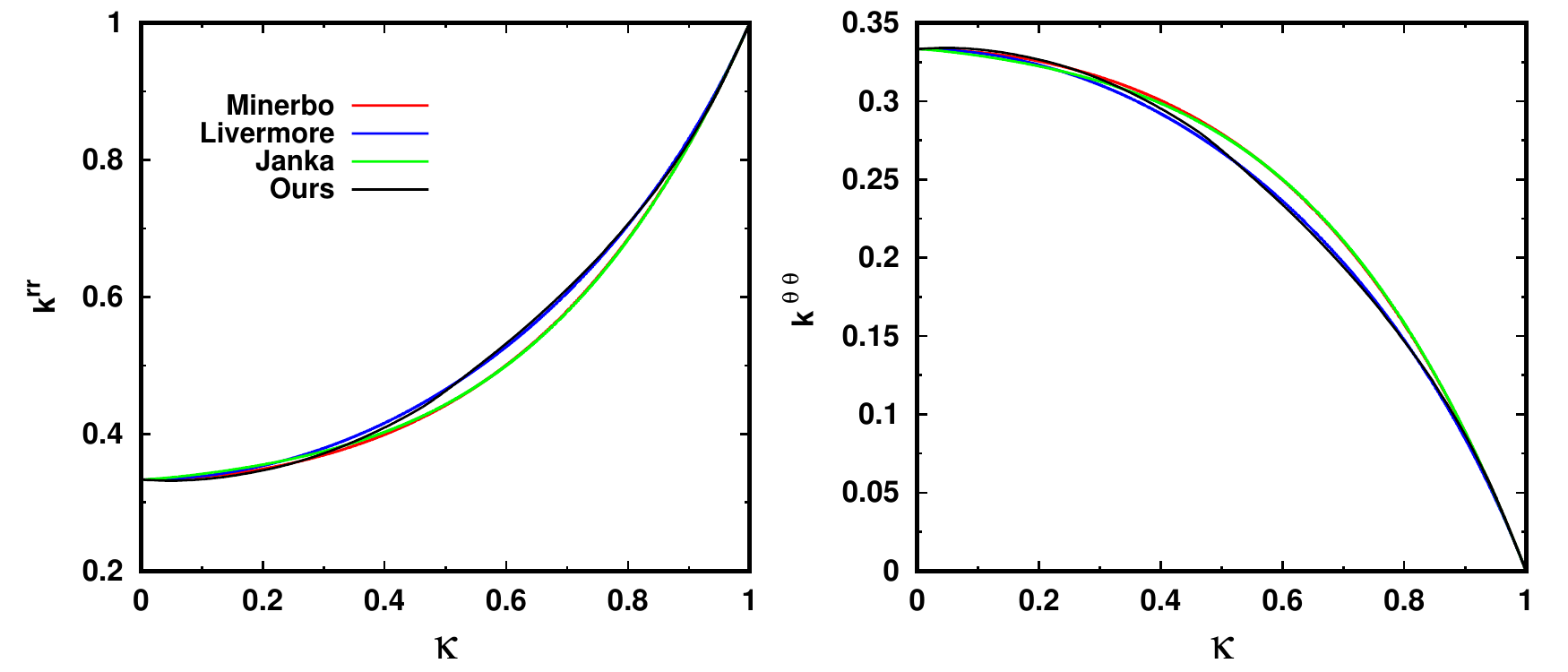}
    \caption{Non-trivial components of Eddington tensor ($k^{ij}$) computed from angular distributions of $f_{\rm n}$ obtained by our method, which is displayed as a function of $\kappa$ with a black line in each panel. Left: $k^{rr}$. Right: $k^{\theta \theta}$. We note that $k^{\phi \phi}$ is the same as $k^{\theta \theta}$ in cases with spherically symmetric radiation field. For comparisons, we display those computed from three representative analytical closure relations. Red: Minerbo. Blue: Livermore. Light-green: Janka.}
    \label{fig:graph_compare_closure}
\end{figure*}

It is interesting to compare the non-trivial components of Eddington tensor computed from the reconstructed $f_{\rm n}$ to those given by some analytical closure relations. We select three representative ones: Minerbo \cite{1978JQSRT..20..541M}, Livermore \cite{1984JQSRT..31..149L}, and Janka \cite{1991ntts.book.....J}, see also \cite{2015MNRAS.453.3386J} for their explicit forms. We show the result of the comparison in Fig.~\ref{fig:graph_compare_closure}. As shown in the plots, the Eddington tensor obtained from our method is similar to that given by Minerbo closure relation at $\kappa \lesssim 0.3$, and then it transits to that given by Livermore one with increasing $\kappa$.

It is deserved to be mentioned that the arbitrary rank of moments can be computed from the reconstructed $f_{\rm ax}$, which can be used for computations in neutrino-matter interactions and neutrino advection. For instance, the third moment is mandatory in energy-dependent two-moment neutrino transport, if relativistic effects (such as fluid-velocity dependence) are taken into account. We also note that it is possible to use our data to give a closure relation for multi-D CCSN simulations by carrying out a coordinate transformation of neutrino momentum space (see also next section for more details). More detailed discussions and the demonstrations are postponed in future work.

Before closing this section, let us remark on computing $f_{\rm ax}$ from $f_{\rm n}$. We integrate reconstructed $f_{\rm n}$ over $\mu$-direction (from $-1$ to $1$), i.e., compute the zero-th moment of $f_{\rm n}$. From Eq.~\ref{eq:def_fnorm1}, $f(\mu=1)$ can be computed as,
\begin{equation}
f(\mu=1) =  \frac{ N }{ 2 \pi \int_{-1}^{1} d \mu f_{\rm n}(\mu) },
\label{eq:fnormrecover}
\end{equation}
where $N$ denotes the zero-th moment of $f$. From Eq.~\ref{eq:def_fnorm1}, we can fully recover $f_{\rm ax}$ by using the reconstructed $f_{\rm n}$ and $f(\mu=1)$.

\section{Demonstration}\label{sec:demo}
We demonstrate the capability of our method by applying it to another CCSN model. We employ neutrino data of a 2D CCSN model \cite{2019ApJ...880L..28N}, in which the neutrino transport is determined by solving full Boltzmann equations. We compute the energy-dependent zero-th and first angular moments by using the distribution function of neutrinos ($f$) in the neutrino data. We reconstruct $f_{\rm ax}$ (azimuthal-averaged distributions; see Eq.~\ref{eq:def_fax}) from the zero-th and first moments by using our method\footnote{In this demonstration, we do not repeat the retrieval procedures presented in previous sections, but rather use a table, in which the best-fit parameters are capsulated. As mentioned already, the data-table is publicly available; hence, this test would be a good example of how to use it.}, and then compare the results to the originals.

\begin{figure*}
    \includegraphics[width=\linewidth]{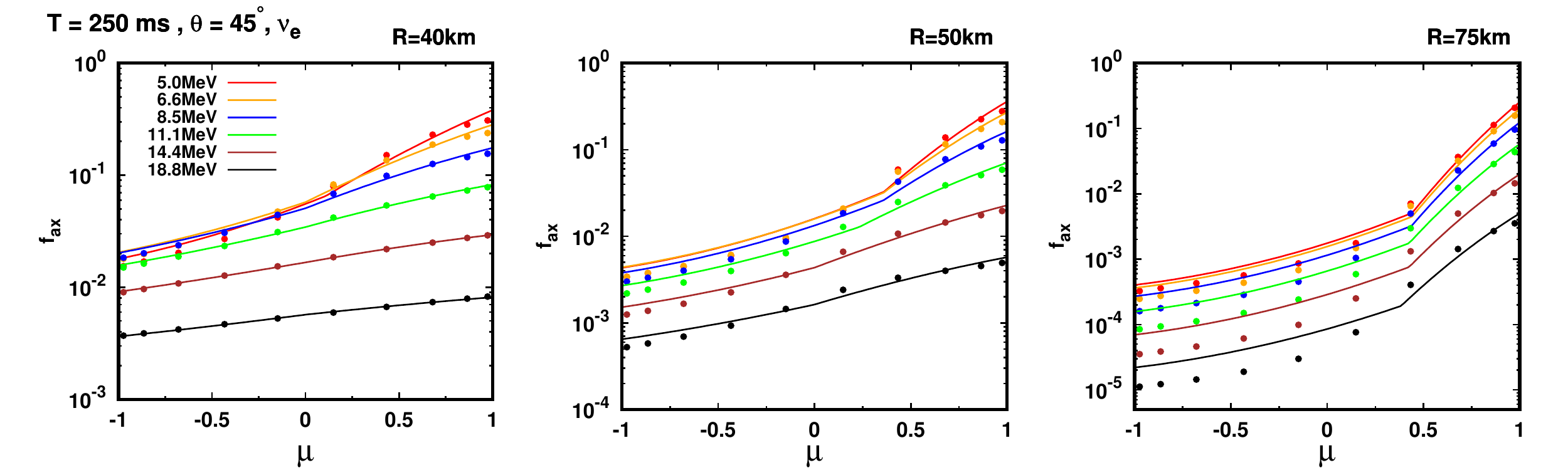}
    \caption{Comparing the azimuthal-averaged angular distributions of neutrinos computed by our method (solid lines) to the originals (our 2D CCSN simulation with Boltzmann neutrino transport) which are displayed with dots. In this comparison, we employ the $\nu_e$ data at the time snapshot at $250$ ms and along a radial ray with $\theta=45^{\circ}$. Color distinguishes the neutrino energy. From left to right, we display the results at different radii, $R=40, 50,$ and $75$ km, respectively.}
    \label{fig:graph_2Denedepe_nue_angdist_reconst_vs_Boltz_th45}
\end{figure*}

There are two remarks before moving on the detail of the demonstration. First, one may wonder if this test is not necessary, since the capability of our method has been already tested by comparing the reconstructed $f_{\rm ax}$ to that obtained from a spherically symmetric Boltzmann simulation (see Fig.~\ref{fig:graph_compare_angfn_reconst_vs_Boltz}). It should be noted, however, that all fitting parameters are calibrated with the neutrino data of the same CCSN simulation, indicating that the consistency between the reconstructed $f_{\rm ax}$ and the original may be a trivial result. On the other hand, we employ an independent CCSN model in this demonstration with different input physics (neutrino matter interactions \cite{2019ApJS..240...38N} and equation-of-state \cite{2017JPhG...44i4001F}) and, more importantly, this is a multi-D model; hence, we can fairly asses the capability of our method. Next, let us remark on a treatment of flux factor. In general, it is defined by the norm of the first angular moment divided by the zero-th one. In this demonstration, however, we adopt the radial component of the first moment. The main reason of the simplicity is that our method is developed based on a spherically symmetric CCSN model, indicating that it is not capable of capturing the angular structure associated with non-radial fluxes. 

\begin{figure*}
    \includegraphics[width=\linewidth]{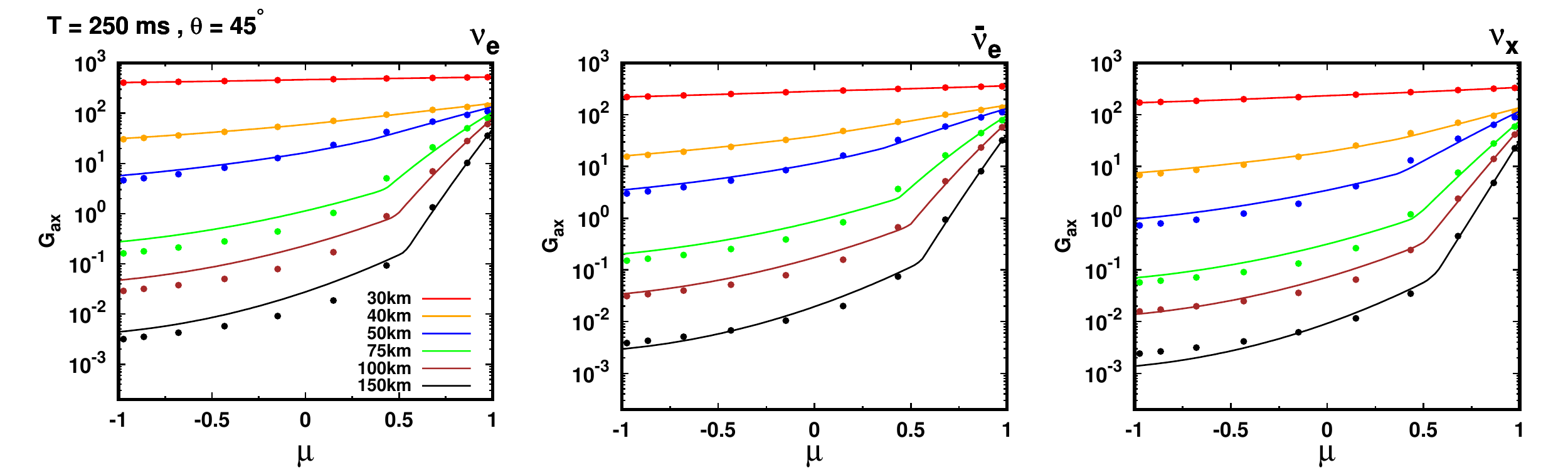}
    \caption{Comparing energy-integrated angular distributions of neutrinos ($G_{\rm ax}$). In the computation of $G_{\rm ax}$, we first reconstruct $f_{\rm ax}$ at each energy by using our method, and then summed over the energy range of $0-300$ MeV (see Eq.~\ref{eq:def_G} and text for more details.). We display them with solid lines and the dots denote the originals (those obtained from the 2D Boltzmann simulation). We employ the neutrino data at $250$ ms and along a radial ray with $\theta=45^{\circ}$. Color distinguishes the radius. From left to right, we show the results for $\nu_e$, $\bar{\nu}_e,$ and $\nu_x$, respectively.}
    \label{fig:graph_2Deneinteg_angdist_reconst_vs_Boltz_th45}
\end{figure*}

\begin{figure*}
    \includegraphics[width=\linewidth]{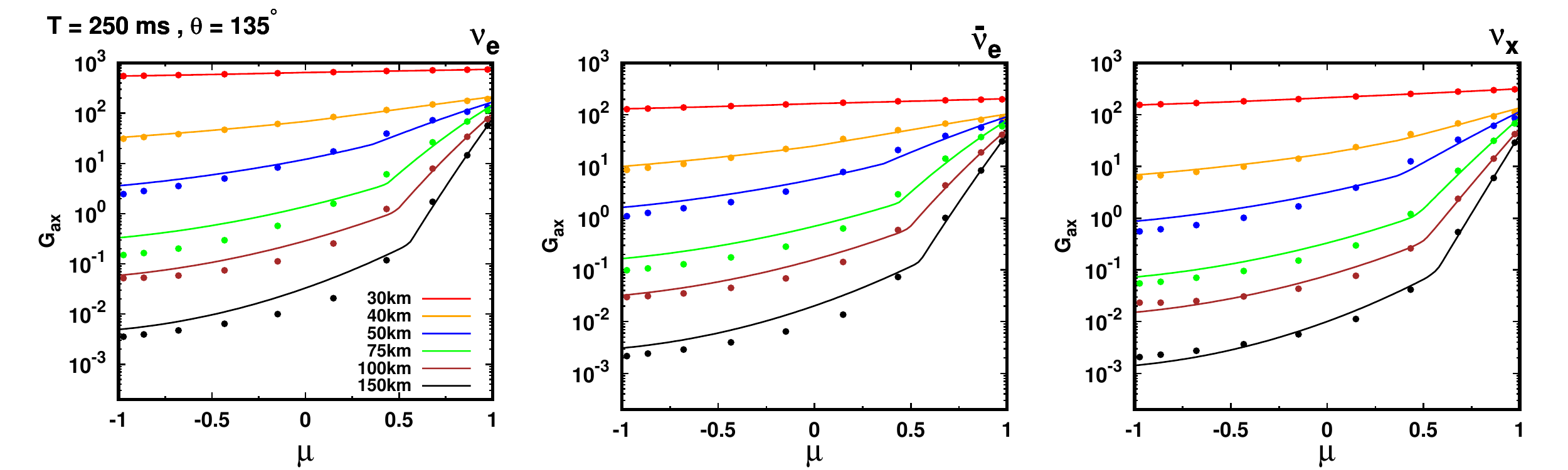}
    \caption{Same as Fig.~\ref{fig:graph_2Deneinteg_angdist_reconst_vs_Boltz_th45} but for a different radial ray: $\theta=135^{\circ}$.}
    \label{fig:graph_2Deneinteg_angdist_reconst_vs_Boltz_th135}
\end{figure*}

We note, however, that there may be a better prescription regarding the treatment of the flux factor in our method. We adopt the flux factor (following the definition, i.e., including non-radial components) to reconstruct $f_{\rm ax}$. The obtained angular distribution can be interpreted as the azimuthal-averaged distribution with respect to the direction of the neutrino flux. In other words, the non-trivial (diagonal) components of Eddington tensor obtained by our method can be interpreted as the eigen-values of the two-rank matrix\footnote{We note that the Eddington tensor can be always diagonalized, since it is a symmetric matrix. See also \cite{2020ApJ...903...82I} for the detailed mathematical property of the Eddington tensor.}. The angular distributions of neutrinos expressed with angles measured from the radial direction can be obtained by a coordinate transformation in neutrino momentum space. The obtained $f_{\rm ax}$ through the coordinate transformation seems to be more realistic than that computed only by the radial component of the neutrino flux, since the condition of axial symmetry in neutrino momentum space would be a reasonable approximation with respect to the direction of the flux. This treatment may be useful, in particular for determining a closure relation in multi-D two moment neutrino transport. It is interesting to see how the CCSN dynamics is sensitive to the different treatment of flux factor in our method, although the detailed study is beyond the scope of this paper and will be addressed in future work.

In the demonstration, we employ a neutrino data at the time snapshot of $250$ ms after bounce. In this time snapshot, we found strongly asymmetric $\nu_e$ and $\bar{\nu}_e$ emissions associated with PNS kick \cite{2019ApJ...880L..28N}, which triggers the occurrence of ELN-crossings at $R \gtrsim 50$ km in the northern hemisphere (see \cite{2019ApJ...886..139N} for more details). The neutrino radiation field is, hence, convenient to assess the capability of our method for searching ELN-crossings, which is one of the items to be considered in this paper.

Fig.~\ref{fig:graph_2Denedepe_nue_angdist_reconst_vs_Boltz_th45} portrays the energy-dependent features of angular distributions of reconstructed $f_{\rm ax}$, focusing on $\nu_e$. We display the results at different radii ($40, 50,$ and $75$ km from left to right panel) along the same radial ray with $\theta=45^{\circ}$. As shown in the figure, the reconstructed $f_{\rm ax}$ has a similar profile as that in the original data (obtained from Boltzmann simulation) at $R \lesssim 50$ km. On the other hand, we find a relatively large systematic errors at $R = 75$ km, in particular at the region of $\mu < 0.5$. This trend is consistent with our previous argument discussed in Sec.~\ref{sec:Construction}, saying that our method is not capable of capturing the angular profile for incoming neutrinos accurately in cases with forward-peaked distributions (the flux factor for $10$ MeV neutrinos is roughly $0.8$ at $75$ km). We also find that other effects (multi-D neutrino transport and different input physics) seem to slightly enhance the deviation (see also Fig.~\ref{fig:graph_compare_angfn_reconst_vs_Boltz} for the case of $\kappa=0.8$).

It should be stressed, however, that the reconstructed $f_{\rm ax}$ captures the essential profile for outgoing neutrinos regardless of neutrino energy and spatial positions. To see the trend more comprehensively, we check them for other neutrino species and at other spatial locations. To extract the essence, we compute the energy-integrated angular distribution ($G_{\rm ax}$), which is defined as
\begin{equation}
G_{\rm ax}(\mu) = \int d (\frac{\varepsilon^3}{3}) f_{\rm ax}(\varepsilon,\mu),
\label{eq:def_G}
\end{equation}
where $\varepsilon$ denotes the neutrino energy, and we carry out the $\varepsilon$-integration in the right hand side of Eq.~\ref{eq:def_G} with the unit of MeV. Following the definition, we compute $G_{\rm ax}$ from the reconstructed $f_{\rm ax}$ at different energies. Figs.~\ref{fig:graph_2Deneinteg_angdist_reconst_vs_Boltz_th45}~and~\ref{fig:graph_2Deneinteg_angdist_reconst_vs_Boltz_th135} portray the results along the radial ray with $\theta=45^{\circ}$ and $135^{\circ}$, respectively. We confirm that the reconstructed angular distributions of $G_{\rm ax}(\mu)$ at $\mu \gtrsim 0.5$ are good agreement with those obtained from Boltzmann for all neutrino species and even in cases with strongly forward-peaked angular distributions\footnote{For instance, the flux factor at $R=150$ km for the average energy of neutrinos ($\sim 10$ MeV) is larger than $0.9$, indicating that the angular distribution is strongly forward-peaked.}. The accurate reconstruction of $f_{\rm ax}$ for outgoing neutrinos is one of the advantages in our method.

\begin{figure*}
    \includegraphics[width=\linewidth]{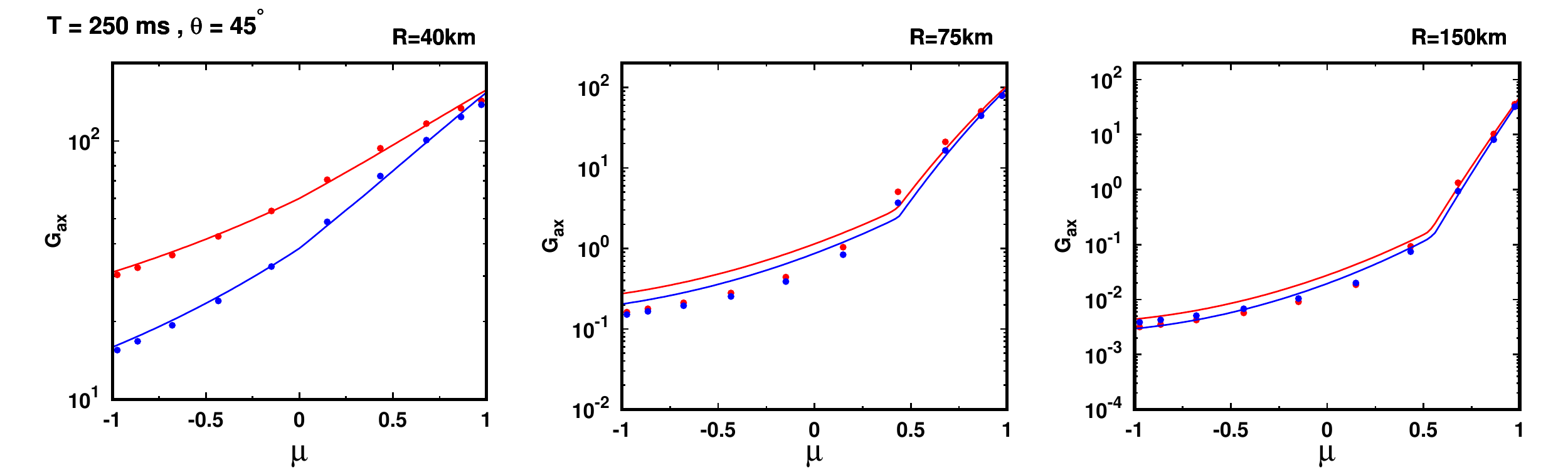}
    \caption{Reconstructed energy-integrated angular distributions ($G_{\rm ax}$) of $\nu_e$ (red) and $\bar{\nu}_e$ (blue). The solid lines are those obtained from our method, and the dots correspond to the results of 2D Boltzmann simulation. We employ the neutrino data at $250$ ms and along a radial ray with $\theta=45^{\circ}$. From left to right, we display the results at different radii, $R=40, 75,$ and $150$ km, respectively. This comparison is useful to assess the capability of ELN-crossing searches based on two-moment methods; see text for more details.}
    \label{fig:graph_2Deneinteg_angdist_reconst_vs_Boltz_th45_ELN}
\end{figure*}

\begin{figure*}
    \includegraphics[width=\linewidth]{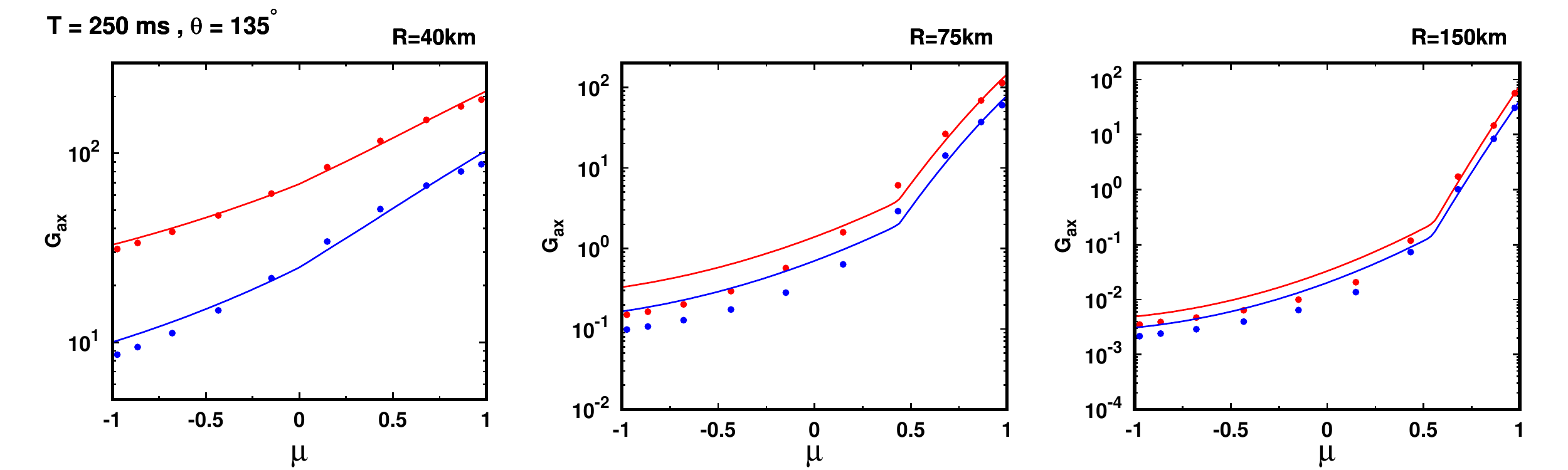}
    \caption{Same as Fig.~\ref{fig:graph_2Deneinteg_angdist_reconst_vs_Boltz_th45_ELN} but for a different radial ray: $\theta=135^{\circ}$.}
    \label{fig:graph_2Deneinteg_angdist_reconst_vs_Boltz_th135_ELN}
\end{figure*}

Finally, let us discuss the reliability of approximate ELN-crossing searches based on two moment neutrino transport. In Figs.~\ref{fig:graph_2Deneinteg_angdist_reconst_vs_Boltz_th45_ELN}~and~\ref{fig:graph_2Deneinteg_angdist_reconst_vs_Boltz_th135_ELN}, we display the angular distribution of $G_{\rm ax}$ for $\nu_e$ and $\bar{\nu}_e$ in the same panel. In the former figure, we select a radial ray with $\theta=45^{\circ}$, in which ELN-crossings are observed in the original neutrino distribution at $R \gtrsim 50$ km. In the latter figure, on the other hand, we select a radial ray with $\theta=135^{\circ}$, in which no ELN-crossings occur in the entire post-shock region. At $R=40$ km (left panel) where ELN crossings are not observed in the CCSN model at both $\theta=45^{\circ}$ and $135^{\circ}$, we confirm that the systematic error of reconstructed $G_{\rm ax}(\mu)$ is very small; consequently, the reconstructed $G_{\rm ax}(\mu)$ of $\nu_e$ and $\bar{\nu}_e$ do not cross each other, which is consistent with that in the originals. We should mention a caveat, however, that $G_{\rm ax}$ at $\mu=1$ for $\nu_e$ and $\bar{\nu}_e$ is very close each other for $\theta=45^{\circ}$ at $40$ km, indicating that the small systematic error potentially change the result. On the contrary, our method provides a robust assessment for ELN crossing at $\theta=135^{\circ}$ (and $40$ km), since the difference of angular distributions of $G_{\rm ax}$ between $\nu_e$ and $\bar{\nu}_e$ is much larger than the systematic errors of the reconstruction.

At $R=75$ km and $150$ km, on the other hand, we found a fatal issue; the retrieved angular distributions of $G_{\rm ax}$ of $\nu_e$ and $\bar{\nu}_e$ at $\theta=45^{\circ}$ do not cross each other, whereas those in the original data do (see the middle and right panels in Fig.~\ref{fig:graph_2Deneinteg_angdist_reconst_vs_Boltz_th45_ELN}), indicating that our method fails to judge the ELN-crossing. The misjudgement is attributed to the fact that the systematic errors overwhelm the actual difference between $\nu_e$ and $\bar{\nu}_e$ angular distributions. It should be stressed that the ELN-crossing in the original data is so tiny. We note that such tiny crossings are common in CCSN environment (see, e.g., \cite{2019ApJ...886..139N,2020PhRvD.101d3016A}), indicating that the small error does critical harm to the judgement. We conclude that ELN-crossing searches by our method may be valid only at $r \lesssim 50$ km or $\kappa \lesssim 0.5$ for the average energy of neutrinos.

\begin{figure*}
    \includegraphics[width=\linewidth]{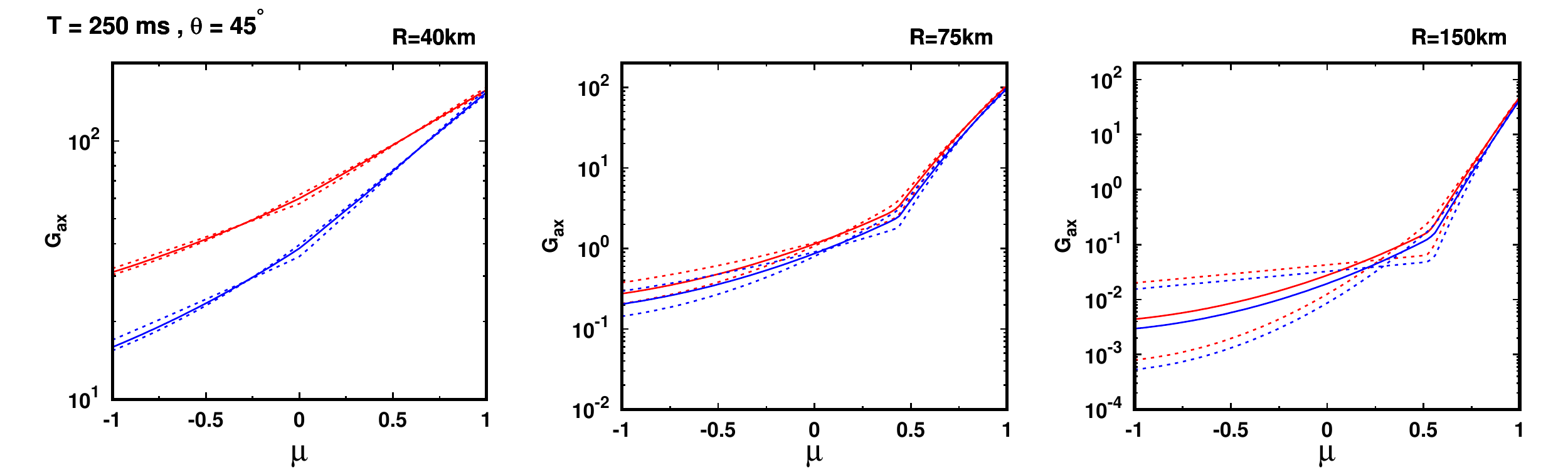}
    \caption{Same as Fig.~\ref{fig:graph_2Deneinteg_angdist_reconst_vs_Boltz_th45_ELN} but we add the results with different set of fitting parameters, in which the $f_{\rm n}$ at $\mu=-1$ is changed in the range of dispersion (see text for more details). They are displayed with dotted lines, meanwhile the solid line represents the results with best-fit parameters. As shown in these panels, there are large dispersion of the angular distribution at $\mu \lesssim 0$, which increases with radius.}
    \label{fig:graph_2Deneinteg_angdist_reconst_th45_ELN_Error}
\end{figure*}

\begin{figure*}
    \includegraphics[width=\linewidth]{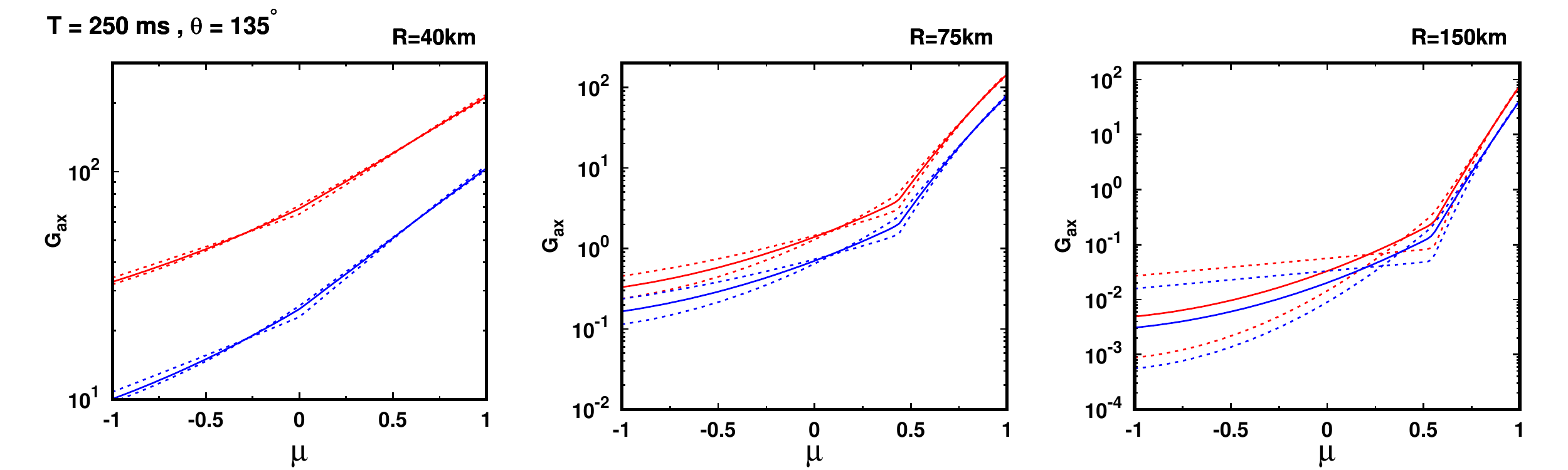}
    \caption{Same as Fig.~\ref{fig:graph_2Deneinteg_angdist_reconst_th45_ELN_Error} but for $\theta=135^{\circ}$.}
    \label{fig:graph_2Deneinteg_angdist_reconst_th135_ELN_Error}
\end{figure*}

To strengthen our statement regarding the limitation of ELN-crossing searches, we reconstruct $G_{\rm ax}(\mu)$ of $\nu_e$ and $\bar{\nu}_e$ from different values of parameters in our method. We change the fitting coefficients for the correlation between $f_{\rm n}(\mu=-1)$ and $\kappa$ (see Sec.~\ref{sec:Correlations}). We consider the two cases that represent the dispersion of the correlation (see red and blue lines in Fig.~\ref{fig:graph_corre_kappa_fnmin1}). With replacing the fitting function, we take the same procedure to determine seven free parameters to obtain $f_{\rm n}$ as described in Sec.~\ref{sec:Construction}. The results of retrieved $G_{\rm ax}$ are displayed as dashed lines in Figs.~\ref{fig:graph_2Deneinteg_angdist_reconst_th45_ELN_Error}~and~\ref{fig:graph_2Deneinteg_angdist_reconst_th135_ELN_Error} (we also show the result of best-fit parameters with solid lines in the panel as a reference). As expected, the dispersion of angular distribution of $G_{\rm ax}$ at $40$ km is very small, indicating that our method provides a robust judgement for ELN-crossings. On the contrary, there are large uncertainties of $G_{\rm ax}$ at $\mu=-1$ at $75$ km and $150$ km, which makes the sign of ELN indeterminate. This is an evidence that the judgement of ELN-crossing in our method is not reliable at large radii.

The above argument would illustrate a common issue for other approximate ELN-crossing searches based on two-moment neutrino transport. It also casts doubt on the conclusion in \cite{2020arXiv201206594A,2021PhRvD.103f3013C}. In these studies, they judge the occurrence of ELN-crossing if there exists $\mathcal{F}$ fulfilling the condition that the product of $I_0 I_F$ becomes negative (where $I_0$ and $I_F$ denote the ELN and that integrated with $\mathcal{F}$, see \cite{2020JCAP...05..027A} for more details); in other words, they judge no ELN crossings only if arbitrary $\mathcal{F}$ provides $I_0 I_F > 0$. We reckon that this criterion would lead to an optimistic judgement with respect to the occurrence of ELN-crossing, and there is a danger that they capture the spurious crossings\footnote{One may think that the spurious crossings never occur in their method as long as $\mathcal{F}$ is positive (see also \cite{2020JCAP...05..027A}). However, the statement is only true if the full angular moments of neutrinos are considered. If we truncate the angular moments at a certain rank (as we do in the two-moment method), it potentially generates spurious crossings.}. In fact, our demonstration suggests that it is easy to generate ELN-crossings artificially at $R \gtrsim 50$km by changing $f_{\rm n}$ at $\mu=-1$ within a certain range of the dispersion (see in Figs.~\ref{fig:graph_2Deneinteg_angdist_reconst_th45_ELN_Error}~and~\ref{fig:graph_2Deneinteg_angdist_reconst_th135_ELN_Error}). This fact urges reconsideration of their conclusions, and it would be necessary to make detailed comparisons with the results of multi-angle neutrino transport to substantiate their claim.

\section{Summary}\label{sec:summary}

In this paper, we develop a novel method by which to construct neutrino angular distribution in momentum space from the zero-th and first angular moments. In our method, we employ two quadratic functions in a piecewise fashion to determine the angular ($\mu$-) distribution, and the seven free-parameters are determined by referring the neutrino data of a spherically symmetric CCSN simulation with full Boltzmann neutrino transport \cite{2017ApJ...847..133R}. To narrow down the parameter-space, we search correlated quantities in the angular distribution of neutrinos to a flux factor ($\kappa$). By using the results of correlation, we determine the best parameters that minimize differences between reconstructed angular distribution of neutrinos and those of the original. We extend our method with a makeshift but appropriate prescription to cover the cases with strong forward-peaked angular distribution ($\kappa \gtrsim 0.9$) where the Boltzmann simulation can not provide the accurate data. Consequently, we determine the all of the free parameters to arbitrary $\kappa$; i.e., we complete the construction of a data-table capsulating all best-fit parameters. We note that other useful quantities (such as non-trivial components of Eddington tensor) are also included in the file.

By using our method, we demonstrate reconstruction of $f_{\rm ax}$ by using the neutrino data from one of the most recent 2D CCSN model with full Boltzmann neutrino transport \cite{2019ApJ...880L..28N}. We assess the capability of our method by comparing the reconstructed $f_{\rm ax}$ to the originals. The demonstration lends confidence to our method; indeed, essential features of the angular distributions, in particular for outgoing neutrinos, can be well reconstructed, which is the strongness in our method. On the other hand, we also underline the weakness in our method; the large systematic errors emerge in the angular distributions for incoming neutrinos, and it increases with $\kappa$. This issue is critical for ELN-crossing searches; indeed, our method fails to capture the crossings at $\kappa \gtrsim 0.5$ (or $R \gtrsim 50$ km) for the CCSN model. We also find that all approximate ELN-crossing searches based on two moment neutrino transport would miss the judgement at $\kappa \gtrsim 0.5$, since the systematic errors in the reconstructed $f_{\rm ax}$ would overwhelm the actual difference of angular distribution between $\nu_e$ and $\bar{\nu}_e$; indeed, ELN-crossings in CCSN environments are usually very tiny. Our result suggests that the jugement of ELN-crossings based on other approximate prescriptions with two moment neutrino transport is not accurate at $R \gtrsim 50$ km or $\kappa \gtrsim 0.5$. It should also noted that we assess the applicability of ELN-crossing searches with two-moment methods by using a different approach (Lucas Johns and Hiroki Nagakura in prep), and the conclusion is consistent with our finding in this paper.

We note that this study gives a hint of how to improve the accuracy of ELN-crossing searches with approximate neutrino transport. As described above, the biggest uncertainty in our method is to construct angular distributions for incoming neutrinos. This indicates that the accuracy of ELN-crossing may be substantially improved if we can correct them appropriately. We reckon that the ray-tracing method may be convenient for the purpose. The essential idea is as follows. After obtaining $f_{\rm ax}$ by using the present method, we perform a ray-tracing computation to determine $f(\mu=-1)$, i.e., we solve geodesic equations for incoming neutrinos along each radial ray\footnote{In our method, we only solve the geodesic equation along the radial ray of different solid angle, which is much computationally cheaper than full Boltzmann neutrino transport, indicating that it would be feasible.}. In the ray-tracing simulation, we need to compute the reaction kernels of neutrino-matter interactions, that can be preliminary evaluated by using the reconstructed $f_{\rm ax}$. Once the ray-tracing computation is over, we then replace $f(\mu=-1)$ and assess the ELN-crossing by using $G_{\rm ax}(\mu=1)$ and the renewed $G_{\rm ax}(\mu=-1)$ (see Eq.~\ref{eq:def_G}). We note that the occurrence of ELN crossing can be approximately judged by checking only the sign of ELN at $\mu=-1$ and $1$ in most of the situation. In fact, the number of ELN crossings found in previous CCSN simulations is usually one (although there may be exceptions \cite{2021PhRvD.103f3013C}), indicating that the two signs become opposite each other in this case. This study is currently underway, and the results will be reported in our forthcoming paper.

Let us close this paper with giving a recipe of how to construct angular distributions of neutrinos from the zero-th and first angular moments with using the data-table that capsulates all necessary items for the computation.

\begin{enumerate}
\item Compute flux factors ($\kappa$) of neutrinos from the zero-th and first angular moments. If the neutrino transport is solved with energy-dependent schemes, it would be preferred to compute $\kappa$ at each neutrino energy\footnote{As a reference, we refer the readers to see Sec.~\ref{sec:demo}: how our method can be applied in multi-D CCSN model.}. It may be useful to compute $\kappa$ from the energy-integrated (or average-energy) moments\footnote{This is essential if the neutrino transport is solved under the gray approximation.}, although the reconstructed angular distribution may be less accurate than that integrated from the energy-dependent distributions. We note that there are two options to compute $\kappa$ in the energy-integrated case; one is computed from the number density (zero-th moment) and number flux (first moment), and the other is computed from energy density (zero-th moment) and energy flux (first moment). The choice depends on the purpose of the analysis; for instance, the former choice (number density and flux) would be appropriate for ELN-crossing searches.
\item Extract the seven parameters characterizing the normalized angular distribution of $f_{\rm n}$ (see Eq.~\ref{eq:def_fnorm1} for the definition of $f_{\rm n}$) from the data-table. See also Fig.~\ref{fig:ReconstructBaseForm} for the definition of each parameter. We note that $\kappa$ listed in the table is discretized. This implies that it is necessary to carry out an interpolation so as to construct an angular distribution matching with the given $\kappa$\footnote{We utilize a fine mesh of $\kappa$ in the table; hence, it would be no problem to adopt the parameters at the nearest cell center.}. Let us provide an example. We take the free parameters at neighboring $\kappa$-cells on both sides of the given $\kappa$, and then reconstruct their angular distributions of $f_{\rm n}$. We then obtain $f_{\rm n}$ at given $\kappa$ by interpolating linearly between the two distributions at each angular point ($\mu$). We note that $\int d\mu f_{\rm n}$ is also listed in the data-table; hence, this value should be also interpolated, which will be used in the next step.
\item By using the $N$ (the zero-th angular moment of $f$) and $\int d\mu f_{\rm n}$, we compute $f(\mu=1)$ from Eq.~\ref{eq:fnormrecover}.
\item We then compute the azimuthal-averaged angular distribution of neutrinos ($f_{\rm ax}$) by using the obtained $f(\mu=1)$ and $f_{\rm n}$ from Eq.~\ref{eq:def_fnorm1}.
\end{enumerate}

\section{Acknowledgments}
We are grateful to Sherwood Richers for useful comments and discussions. We also acknowledge conversations with Sam Flynn, Nicole Ford, Evan Grohs, Jim Kneller, Gail McLaughlin, Don Willcox, Taiki Morinaga, and Eirik Endeve. H.N acknowledges support from the U.S. Department of Energy Office of Science and the Office of Advanced Scientific Computing Research via the Scientific Discovery through Advanced Computing (SciDAC4) program and Grant DE-SC0018297 (subaward 00009650). The numerical computations of our CCSN models were performed on the K computer at the RIKEN under HPCI Strategic Program of Japanese MEXT (Project ID: hpci 160071, 160211, 170230, 170031, 170304, hp180179, hp180111, and hp180239). L.J. acknowledges support provided by NASA through the NASA Hubble Fellowship grant number HST-HF2-51461.001-A awarded by the Space Telescope Science Institute, which is operated by the Association of Universities for Research in Astronomy, Incorporated, under NASA contract NAS5-26555.

\bibliography{bibfile}


\end{document}